\documentclass[final,authoryear,3p,times,twocolumn]{elsarticle}
\usepackage{graphics}
\usepackage{amssymb}

\journal{BioSystems}

\begin{document}
\begin{frontmatter}

\title{Coevolutionary games -- a mini review}

\author[Perc]{Matja{\v z} Perc\corref{mp}}
\cortext[mp]{matjaz.perc@uni-mb.si}
\author[Szolnoki]{Attila Szolnoki\corref{as}}
\cortext[as]{szolnoki@mfa.kfki.hu}

\address[Perc]{Department of Physics, Faculty of Natural Sciences and Mathematics, University of Maribor, Koro{\v s}ka cesta 160, SI-2000 Maribor, Slovenia}
\address[Szolnoki]{Research Institute for Technical Physics and Materials Science, P.O. Box 49, H-1525 Budapest, Hungary}

\begin{abstract}
Prevalence of cooperation within groups of selfish individuals is puzzling in that it contradicts with the basic premise of natural selection. Favoring players with higher fitness, the latter is key for understanding the challenges faced by cooperators when competing with defectors. Evolutionary game theory provides a competent theoretical framework for addressing the subtleties of cooperation in such situations, which are known as social dilemmas. Recent advances point towards the fact that the evolution of strategies alone may be insufficient to fully exploit the benefits offered by cooperative behavior. Indeed, while spatial structure and heterogeneity, for example, have been recognized as potent promoters of cooperation, coevolutionary rules can extend the potentials of such entities further, and even more importantly, lead to the understanding of their emergence. The introduction of coevolutionary rules to evolutionary games implies, that besides the evolution of strategies, another property may simultaneously be subject to evolution as well. Coevolutionary rules may affect the interaction network, the reproduction capability of players, their reputation, mobility or age. Here we review recent works on evolutionary games incorporating coevolutionary rules, as well as give a didactic description of potential pitfalls and misconceptions associated with the subject. In addition, we briefly outline directions for future research that we feel are promising, thereby particularly focusing on dynamical effects of coevolutionary rules on the evolution of cooperation, which are still widely open to research and thus hold promise of exciting new discoveries.
\end{abstract}


\end{frontmatter}

\section{Introduction}
\label{intro}

Cooperation and defection are the two strategies that are usually at the heart of every social dilemma \citep{dawesXarp80}. While cooperative individuals contribute to the collective welfare at a personal cost, defectors choose not to. Due to the resulting lower individual fitness of cooperators the selection pressure acts in favor of the defectors, thus designating the evolution of cooperation as a dilemma standing on its own. Established by \citet{maynardXn73}, evolutionary game theory \citep{maynardX82, weibullX95, gintisX00, nowakX06} provides a competent theoretical framework to address the subtleties of cooperation among selfish and unrelated individuals. The prisoner's dilemma game in particular, is considered a paradigm for tackling the problem of cooperation \citep{axelrodX84}. The game promises a defecting individual the highest fitness if facing a cooperator. At the same time, the exploited cooperator is worse off than a defector playing with another defector. According to the fundamental principles of Darwinian selection, cooperation extinction is therefore inevitable. This unadorned scenario is actually realized in the well-mixed prisoner's dilemma game, where defectors reign supreme \citep{hofbauerX98}. Relaxing the inevitability of a social downfall constituted by the well-mixed prisoner's dilemma is the snowdrift or hawk-dove game \citep{maynardXn73}, where mutual defection is individually less favorable than a cooperation-defection pair-up. Accordingly, the snowdrift game allows for stable coexistence of cooperators and defectors in well-mixed populations \citep{ptaylorXmb78}. Completing the triplet is the stag-hunt game \citep{skyrmsX04}, which together with the prisoner's dilemma and the snowdrift game, forms the standard set of social dilemmas that is frequently explored in the current literature [see \textit{e.g.} \citet{macyXpnas02, santosXpnas06, szolnokiXepl09, rocaXplr09}]. Compared with the prisoner's dilemma, the stag-hunt game offers more support for cooperative individuals in that the reward for mutual cooperation is higher than the temptation to defect. Still, cooperation in the stag-hunt game is compromised by the fact that mutual defection is individually more beneficial than being an exploited cooperator, as recently highlighted by \citet{pachecoXprsb09}.

An important realization by the pursuit of cooperation in the context of social dilemmas was the fact that the outcome of evolutionary games in structured populations can be very different from the well-mixed case. In a pioneering work, \citet{nowakXn92b} showed that the introduction of spatial structure via nearest neighbor interactions enabled the cooperators to form clusters on the square lattice and so protect themselves against the exploitation by defectors. Following this discovery, the impact of the spatial structure on the evolution of cooperation has been investigated in detail \citep{nowakXijbc93, hubermanXpnas93, nowakXijbc94, lindgrenXpd94, nowakXpnas94, durrettXtpb94, grimXjtb95, killingbackXprslb96, nakamaruXjtb97, szaboXpre98, brauchliXjtb99, szaboXpre00a, tanimotoXbs07, alonsosanzXbs09, newthXbs09}, and the subject has since been reviewed comprehensively on different occasions \citep{hauertXijbc02, doebeliXel05, szaboXpr07, rocaXplr09}. Notably, the theoretical conjecture that spatial structure may promote cooperation, or at least sustain a multitude of competing strategies has been confirmed experimentally \citep{kerrXn02}, but there also exist evidences that spatial structure may not necessarily favor cooperation \citep{hauertXn04}. Since the impact of the spatial structure on the evolution of cooperation depends on the governing social dilemma, and due to the difficulties associated with the payoff rankings in experimental and field work \citep{milinskiXprsb97, turnerXn99}, it is certainly good practice to test new mechanisms aimed at promoting cooperation on different evolutionary games.

The recent shift from evolutionary games on regular grids to evolutionary games on complex networks [for the latter see \textit{e.g.} \citet{albertXrmp02, newmanXsiamr03, dorogovtsevX03, boccalettiXpr06}] can be considered a step towards more realistic conditions. Indeed, the shift is by no means trivial and bears fascinating results, as recently reviewed by \citet{szaboXpr07}. Quite remarkably, scale-free networks \citep{barabasiXs99} turned out to sustain cooperation by all three above-described social dilemmas \citep{santosXprl05, santosXpnas06, santosXprslb06}, owing predominantly to the heterogeneity that characterizes their degree distribution. Following this seminal discovery, several studies have since elaborated on different aspects of cooperation on scale-free networks, as for example its dynamical organization \citep{gomezgardenesXprl07, puschXpre08}, evolution under clustering \citep{assenzaXpre08}, mixing patterns \citep{rongXpre07}, memory \citep{wxwangXpre06} and payoff normalization \citep{santosXjeb06, masudaXprsb07, zxwuXpa07, szolnokiXpa08}, as well as its robustness in general \citep{poncelaXnjp07, xjchenXpla08} and under intentional attack and error \citep{percXnjp09}. The body of literature devoted to the study of evolutionary games on complex network is extensive, aside from the scale-free architecture hosting the prisoner's dilemma \citep{pachecoXaipcp05, ohtsukiXn06, tangXepjb06, yschenXpa07, wbduXpa08, gomezgardenesXjtb08, floriaXpre09, xliXjpa09, dpyangXpa09} and the snowdrift game \citep{wxwangXpre06, khleeXpa08, rocaXepjb09}, covering also small-world \citep{abramsonXpre01, bjkimXpre02, masudaXpla03, tomochiXsn04, santosXpre05, zhongXepl06, tomassiniXpre06, fuXepjb07, xjchenXpre08, hxyangXcpb08}, social as well as other real-world networks \citep{holmeXpre03, liebermanXn05, vukovXpre05, zxwuXpre06, xjchenXpa07, fuXpla07, luthiXpa08, lozanoXploso08, ykliuXcpl09, luthiXbs09}. Notably, the impact of different interaction topologies has also been studied for evolutionary games outside the realm of the above-described social dilemmas. Examples include the rock-paper-scissors game \citep{szaboXjpa04, szollosiXpre08}, the ultimatum game \citep{kupermanXepjb08} or the public goods game \citep{hxyangXpre09}, and indeed many more studies of the latter games on complex network are expected in the near future.

Besides the conditions generated by spatiality and complex interaction networks, many different mechanisms have been identified that can promote or otherwise affect the evolution of cooperation, and we mention them here briefly. Aside from network reciprocity inherent to games on graphs and complex networks, other prominent rules promoting cooperative behavior are kin selection \citep{hamiltonXjtb64a, hamiltonXjtb64b}, direct reciprocity \citep{axelrodXs81, brandtXjtb06, pachecoXjtb08}, indirect reciprocity \citep{nowakXjtb98, nowakXn98, fehrXn02, brandtXjtb04, nowakXn05, tanimotoXbs07b} and group selection \citep{dugatkinXbs96, traulsenXpnas06, traulsenXbmb08}, as recently reviewed in \citep{nowakXs06}. Moreover, voluntary participation \citep{hauertXs02, szaboXpre02d, szaboXprl02, semmannXn03, hauertXc03, szaboXpre04b, zxwuXpre05, hauertXs07, ychenXpre08}, social diversity \citep{percXpre08, santosXn08}, asymmetric influence of links and partner selection \citep{bjkimXpre02, zxwuXcpl06}, heterogeneous teaching activity \citep{szolnokiXepl07, szolnokiXepjb08}, and the impact of long-term learning \citep{swangXploso08} have been suggested as interesting possibilities that may emerge in real-life systems. The necessary overlap between interaction and replacement graphs \citep{ohtsukiXprl07, ohtsukiXjtb07b, zxwuXpre07} has also been recognized as an important agonist in the evolution of cooperation. Furthermore, the importance of time scales in evolutionary dynamics \citep{pachecoXprl06, rocaXprl06, pachecoXjtb06}, the role of finite population size \citep{nowakXn04b, traulsenXprl05, traulsenXpre06}, and the impact of noise and uncertainties on evolution in general \citep{nowakXjmb95, traulsenXprl04, szaboXpre05, percXnjp06a, percXnjp06b, percXepl06, percXnjp06c, vukovXpre06, tanimotoXpre07b, percXpre07, renXpre07, percXnjp07b} have been investigated as well. Very recently, random explorations of strategies \citep{traulsenXpnas09} and simultaneous adoptions of different strategies depending on the opponents \citep{wardilXepl09} have also been identified as potent promoters of cooperation. Some of these mechanisms will be described more accurately in the subsequent sections, but otherwise the reader is referred to the original works for details.

In the focus of this mini review are evolutionary games with coevolutionary rules. Initiated by \citet{zimmermannX01} and by \citet{ebelXaxv02}, and in some sense motivated by then very vibrant advances in network growth and evolution \citep{strogatzXn01, albertXrmp02}, the subject has evolved into a mushrooming avenue of research that offers new ways of ensuring cooperation in situations constituting a social dilemma. Coevolutionary rules constitute a natural upgrade of evolutionary games since in reality not only do the strategies evolve in time, but so does the environment, and indeed many other factors that in turn affect back the outcome of the evolution of strategies. Coevolutionary rules can affect the links players make (or brake) \citep{ebelXaxv02, zimmermannXpre04, zimmermannXpre05, eguiluzXajs05, pachecoXjtb06, pachecoXprl06, santosXploscb06, hanakiXms06, bielyXpd07, wliXpre07, tanimotoXpre07, fuXpa07, szolnokiXepl08, pachecoXjtb08, percXpre08b, xjchenXpre08b, pestelacciXbt08, vansegbroeckXbmceb08, fuXpre08, fuXpre09, akunXbs09, szolnokiXepl09, tanimotoXieee09, tanimotoXpa09, vansegbroeckXprl09, graeserXepl09} (see Section~\ref{interactions}), the size of the network (or population) \citep{renXcm06b, poncelaXploso08, poncelaXnjp09} (see Section~\ref{growth}), the teaching activity (or reproduction capability)
\citep{szolnokiXnjp08, szolnokiXepjb09} (see Section~\ref{teaching}) and mobility of players \citep{majeskiXc99, vainsteinXpre01, vainsteinXjtb07, helbingXacs08, helbingXpnas09, meloniXpre09, drozXepjb09} (see Section~\ref{mobility}), their age \citep{mcnamaraXn08, starkXprl08, starkXacs08, szolnokiXpre09,dpyangXnjp09} (see Section~\ref{aging}), as well as several other factors \citep{kirchkampXjebo99, gintisXjtb03, axelrodXe04, hamiltonimXprsb05, fortXepl08, hatzopoulosXpre08, dingXijmpc09, moyanoXjtb09, scheuringXjtb09, rankinXev09, szaboXepl09} (see Section~\ref{related}) that eventually affect the outcome of the underlying evolutionary game. Although the majority of coevolutionary rules studied so far affects the network architecture and size, it is important to distinguish these studies from previous, partially closely related works where networks also change or evolve in the course of time \citep{caldrelliXjt98, pfeifferXplosb05, holmeXprl06, grossXjrsi08, castellanoXrmp09}; in particularly so, since the term `coevolution' has in the past been used quite frequently and for rather different processes.

In the continuation of this paper we will review recent advances on evolutionary games with coevolutionary rules, affecting, as mentioned above, the interaction network, the reproduction capability of players, their reputation, mobility or age, more thoroughly. Before that, however, we give in Section~\ref{games} a more technical description of the evolutionary games and strategy adoption rules that we will encounter throughout the paper. Following the main body of the review given in Section~\ref{rules}, we conclude our work and give an outlook in Section~\ref{final}.

\section{Evolutionary games}
\label{games}

As noted in the first paragraph of Section~\ref{intro}, the three main social dilemmas involving pairwise interactions are constituted by the prisoner's dilemma game, the snowdrift game and the stag-hunt game. At least one of these three games is employed in the majority of the works we will review below, and hence we give a more accurate description of them in what follows.

\begin{figure}
\begin{center}
\includegraphics[width=7.5cm]{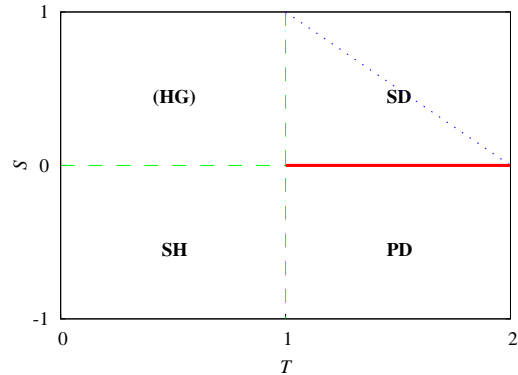}
\caption{Schematic presentation of the two-dimensional $T-S$ parameter plane encompassing the stag-hunt (SH), the prisoner's dilemma (PD) and the snowdrift (SD) game. Borders between games are denoted by dashed green lines. Dotted blue diagonal depicts the $r$-parametrization of the snowdrift game, while the thick red line shows the span of the weak prisoner's dilemma game having $T=b$ as the only main parameter. The upper left quadrant represents the so-called harmony game (HG). The latter, however, does not constitute a social dilemma because there cooperation is always the winning strategy.}
\label{fig:schematic_games}
\end{center}
\end{figure}

Irrespective of which game applies, players can choose either to cooperate or to defect. Notably, other strategies, such as loners [see \textit{e.g.} \citet{hauertXajp05}] or punishers [see \textit{e.g.} \citet{dreberXn08}] are also possible, but their inclusion to evolutionary games with coevolutionary rules has not yet been considered. In general, mutual cooperation yields the reward $R$, mutual defection leads to punishment $P$, and the mixed choice gives the cooperator the sucker's payoff $S$ and the defector the temptation $T$. The standard scaled parametrization entails designating $R=1$ and $P=0$ as fixed, while the remaining two payoffs can occupy $-1 \leq S \leq 1$ and $0 \leq T \leq 2$. Then, if $T>R>P>S$ we have the prisoner's dilemma game, $T>R>S>P$ yields the snowdrift game, and $R>T>P>S$ the stag-hunt game, as schematically depicted in Fig.~\ref{fig:schematic_games}. Without much loss of generality, this parametrization is often further simplified for the prisoner's dilemma game, so that $T=b$ is the only free parameter while $R=1$ and $P=S=0$ are left constant (thick red line in Fig.~\ref{fig:schematic_games}). However, since then the condition $P>S$ is not strictly fulfilled, this version is traditionally referred to as the weak prisoner's dilemma game \citep{nowakXn92b}. An option is also to use $T=b$, $R=b-c$, $P=0$ and $S=-c$, thus strictly adhering to the prisoner's dilemma payoff ranking $T>R>P>S$ while still having a single tunable parameter in the form of the ratio $b/c$. For the snowdrift game one can, in a similar fashion, introduce $r \in [0,1]$ such that $T=1+r$ and $S=1-r$ [see \textit{e.g.} \citet{wxwangXpre06}], thereby again decreasing the effective dimensionality of the parameter space by one. Note also that $r$ characterizes the cost-to-benefit ratio \citep{santosXprl05} and in fact constitutes a diagonal in the snowdrift quadrant of the $T-S$ parameter plane, as shown in Fig.~\ref{fig:schematic_games} by the dotted blue line. It is worth mentioning that other types of parametrization of two-strategy games are possible as well \citep{tanimotoXpre07}, but we focus on the one presented above since it is the most widely used, thus enabling an efficient comparison of different works.

The most frequently employed setup entails that initially each player $x$ is designated either as a cooperator $(s_x=C)$ or defector $(s_x=D)$ with equal probability, and is placed on one of the nodes of the network with degree $k_x$. Evolution of the two strategies is then performed in accordance with a pairwise comparison rule, during which players accumulate their payoffs $\Pi_x$ by playing the game with their neighbors. Subsequently, player $x$ tries to enforce its strategy $s_x$ on player $y$ in accordance with some probability $W(s_x \rightarrow s_y)$ to be specified below. During the simulation procedure the player $x$ and one of its neighbors $y$ are chosen randomly, whereby in accordance with the random sequential update each player is selected once on average during $N$ (network size) such elementary steps, together constituting one full Monte Carlo step \citep{newmanX99}. Alternatively, players can be selected sequentially, albeit this may cause artificial effects. Independently on whether synchronized or the random sequential update is used, however, the time evolution is always discrete. The probability of strategy adoption $W(s_x \rightarrow s_y)$ can be defined in several ways. If the degree $k_x$ of all players is the same and does not change in time, the Fermi function
\begin{equation}
W(s_x \rightarrow s_y)= \frac{1}{1+\exp[(\Pi_y-\Pi_x)/K]}
\label{fermi}
\end{equation}
is a viable option, as proposed by \citet{szaboXpre98}. In Eq.~\ref{fermi} $K$ denotes the amplitude of noise \citep{vukovXpre06, renXpre07}, or equivalently its inverse ($1/K$) the so-called intensity of selection \citep{fudenbergXtpb06, traulsenXjtb07, altrockXnjp09}. In the $K \to 0$ limit player $x$ always succeeds in enforcing its strategy to player $y$ if only $\Pi_x > \Pi_y$ but never otherwise. For $K>0$, however, strategies performing worse may also be adopted based on unpredictable variations in payoffs \citep{percXnjp06a} or errors in the decision making, for example. Importantly, if the degree distribution of the interaction network (note that this is a property that may likely change due to a coevolutionary rule), at any instance of the game, deviates from the case where all players have the same degree, the application of the Fermi function may introduce additional effects since then the impact of the same value of $K$ effectively varies from one player to the other. Indeed, if the degree distribution characterizing the interaction network is heterogeneous, a more successful player (\textit{i.e.} having a larger payoff) can pass its strategy with the probability
\begin{equation}
W(s_x \rightarrow s_y)=(\Pi_x-\Pi_y)/(\Delta \cdot k_q)
\label{pach}
\end{equation}
where $k_q$ is the largest of the two degrees $k_x$ and $k_y$, and $\Delta=T-S$ for the prisoner's dilemma game, $\Delta=T-P$ for the snowdrift game and $\Delta=R-S$ for the stag-hunt game (note that the ranking of payoff elements for each specific game ensures the positive sign of Eq.~\ref{pach}.) Introduced by \citet{santosXprl05}, it is still a popular choice surpassing the difficulties associated with the Fermi function described above, albeit with the downside of being unable to adjust the level of uncertainty by strategy adoptions.

Finally, we mention another frequently used strategy adoption rule in coevolutionary models; namely the so-called richest-following (or `learning from the best') rule \citep{abramsonXpre01, huXpa07, zxwuXpa07}, where the focal player always imitates the strategy of its most successful neighbor \citep{zimmermannXpre04, eguiluzXajs05, wliXpre07, tanimotoXpre07, tanimotoXpa09}. Contrary to the preceding two strategy adoption rules, the richest-following is completely deterministic, in fact exercising the strongest selection between players. Naturally, there also exist other microscopic strategy adoption rules, such as the win-stay-lose-shift rule where the focal player has restricted information on its neighbors, for which the reader is advised to consult the comprehensive review by \citet{szaboXpr07} for more details.

We will use the notation introduced above throughout this work unless explicitly stated otherwise. Also, any deviations with respect to the employed initial setup, simulation procedure or the definition of strategy adoption probability will be noted when applicable.

\section{Coevolutionary rules}
\label{rules}

While it is obvious that strategies of players engaging in evolutionary games evolve in time, the fact that other properties characterizing either their individual attributes or the environment in which the game is staged may simultaneously evolve as well gained foothold only in recent years. Yet the preceding transitions from well-mixed populations to spatial grids and further to complex networks, and in particular their success in explaining the evolution of cooperation, are inviting to further extensions of the theoretical framework, and indeed, the introduction of coevolutionary rules seems like the logical next step. It should need little persuasion to acknowledge that links we make with others change in time, that all of us age, that our roles in life evolve, and that the society we are part of may itself be subject to transformations on a global scale. Coevolutionary rules aim to integrate these processes into the framework of evolutionary games. Perhaps the biggest challenge thereby is, how to do this without directly (or obviously) promoting cooperation. For example, if one introduces a rule that, in the course of time, cooperators should aim to link only with cooperators and defectors only with defectors, it should come as no surprise that such a coevolutionary rule will likely favor the evolution of cooperation. It is demanding, however, to explore and identify successful mechanisms that do not attribute special, not to say fictitious, cognitive skill to players, and do not use a discriminative set of rules for every participating strategy. Thus, coming up with plausible coevolutionary rules is not straightforward, and care must be exercised in order to give both strategies equal credentials. Simply because a strategy is bad for social welfare it should not be assumed that the individuals adopting it are less skilful or sly than their opponents. In fact, rather the opposite seems to apply. For example, defectors should be assumed being just as skilful by selecting appropriate partners as cooperators.

In the following we will review coevolutionary rules affecting the interactions between players (Section~\ref{interactions}), population growth (Section~\ref{growth}), teaching activity (Section~\ref{teaching}), mobility (Section~\ref{mobility}) and aging (Section~\ref{aging}) of players, as well as related aspects (Section~\ref{related}) of individual and global characteristics that may affect strategy dominance in evolutionary games.

\subsection{Dynamical interactions}
\label{interactions}

Coevolutionary rules frequently affect how players link with one another and this section reviews examples thereof. As we have mentioned above, the result of a game with a partner may influence the durability of such a connection. In particular, an unsatisfied player can easily brake a link to look for a more beneficial interaction with another partner. Notably, the network itself does thereby not shrink or grow in size (for the latter see Section~\ref{growth}). Instead, our aim in this subsection is to explore possible rearrangements of an existing network that is driven by the success of players participating in the governing evolutionary game. Since coevolutionary rules affecting the interactions between players were proposed first \citep{zimmermannX01, ebelXaxv02}, the pertaining literature that has accumulated thus far is rather extensive. Works can be partitioned into those that employed strategy independent rules for link adaptations \citep{szolnokiXepl08, percXpre08b, akunXbs09,  tanimotoXieee09, szolnokiXepl09} and those that considered strategies or their performances as factors potentially affecting the rewiring \citep{ebelXaxv02, zimmermannXpre04, zimmermannXpre05, eguiluzXajs05, pachecoXjtb06, pachecoXprl06, santosXploscb06, wliXpre07, fuXpa07, tanimotoXpre07, bielyXpd07, pachecoXjtb08, vansegbroeckXbmceb08, pestelacciXbt08, fuXpre08, xjchenXpre08b, tanimotoXpa09, fuXpre09, vansegbroeckXprl09, qinXpa09}. Notably, the latter distinction is rather crude and sometimes not completely accurate since the rewiring can be performed based on a secondary player property, like reputation \citep{fuXpre08}, attractiveness \citep{xjchenXpre08b} or satisfaction \citep{pestelacciXbt08}, which are typically related with strategy performance over time. It is indeed possible to further distinguish the proposed coevolutionary rules introducing dynamical interactions to those by which the change of the interaction network is driven by the urge to increase the payoff of the focal player directly \citep{ebelXaxv02, santosXploscb06, bielyXpd07, wliXpre07, xjchenXpre08b, pestelacciXbt08, vansegbroeckXbmceb08, tanimotoXpa09, graeserXepl09}, and those by which the rewiring serves also the increase of the payoff but on a global scale, \textit{i.e.} independently of the payoff of the focal player that is affected by the link adaptation \citep{pachecoXjtb06, pachecoXprl06, tanimotoXpre07, fuXpa07, pachecoXjtb08, fuXpre08, fuXpre09, vansegbroeckXprl09}. In the latter case it is thus not necessary to calculate the players payoff prior to rewiring because solely its strategy determines the `life' of a link.

\begin{figure*} [!ht]
\begin{center}
\includegraphics[width=15cm]{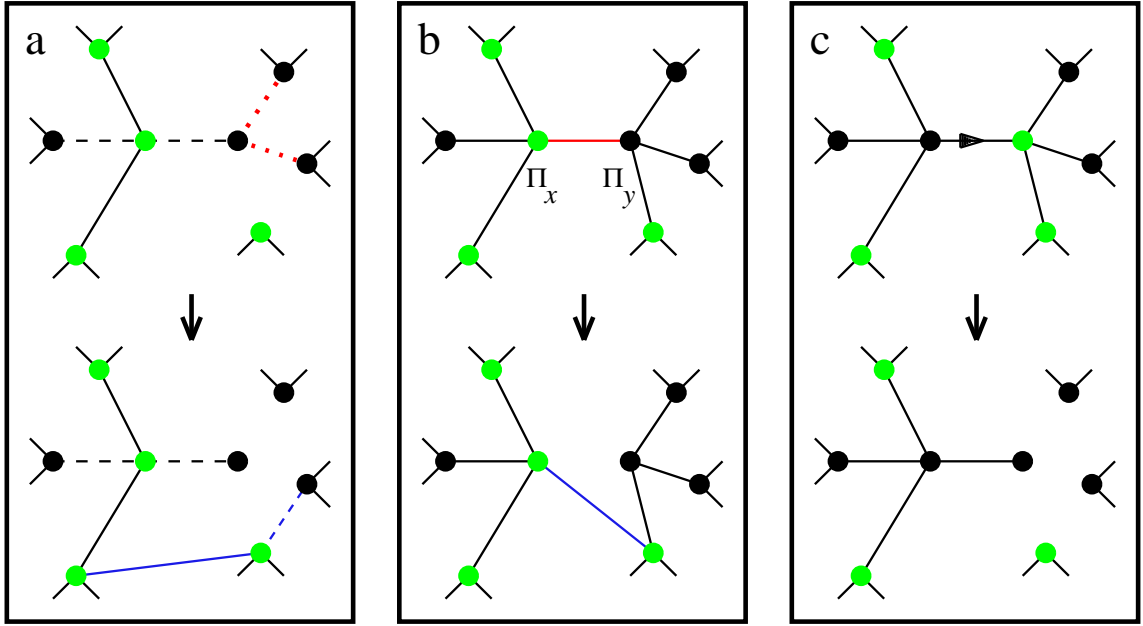}
\caption{Comparative plots of representative coevolutionary rules affecting the interactions between players. In all panels cooperators (defectors) are denoted by green (black) circles. Type A [panel (a)]: By `active linking' the probability to create or to delete a link depends only on its type ($C-C$ links are marked by solid, $C-D$ links by dashed, and $D-D$ links by dotted lines). Links to be deleted (created) are marked by red (blue) color. Type B [panel (b)]: Adverse ties are deleted depending on the payoffs collected by the players having opposite interests. Subsequently, the new link is connected to one of the neighbors of the defeated player. Type C [panel (c)]: Each successful strategy adoption, denoted by a full arrow, evokes the deletion of links of the invaded player, except from the one with the `donor' of the new strategy.}
\label{fig:class_of_updates}
\end{center}
\end{figure*}

Summarizing the above, a simplified but useful classification of interaction-updating rules is presented in Fig.~\ref{fig:class_of_updates}. As suggested in the works mentioned last in the preceding paragraph, the lifetime of a link may depend primarily on the strategies of the players that are connected with it (type A). From this point of view it is straightforward to establish that defector-defector links are short-lived if compared to cooperator-cooperator links since the former are not beneficial for neither of the two involved players, while the later yield mutual gains for both. The second set of coevolutionary rules evaluates the payoffs originating from the investigated link prior to its potential deletion, while the actual removal takes place only if a new neighbor may yield higher benefits (type B) \citep{vansegbroeckXprl09}. And finally, the third set of coevolutionary rules considers the strategy adoption process as pivotal for deciding which links to delete and which to keep (type C). An example thereof is that the invaded player looses all its links except the one with the donor of the new strategy \citep{szolnokiXnjp09, szolnokiXepl09}, as depicted in Fig.~\ref{fig:class_of_updates}(c). There are several real-life situations that can be modeled by the latter rule. From a biological viewpoint, the coevolutionary rule can be linked with an invasion of the subordinate species and the subsequent replacement by a newborn of the victor. A similar phenomenon can be observed in human societies when one changes a job. Typically then the links to former coworkers fade and eventually brake, and new ties are formed primarily with the coworkers from the new working place. Notably, it falls within the same logical set of rules if the player that has successfully passed its strategy is allowed to increase the number of neighbors that are directly connected to it, as was proposed by \citet{szolnokiXepl08}. It should not be overlooked, however, that the strategy adoption process, triggering the deletion and/or addition of links, is itself inherently routed in the payoff difference of the considered players.

An important feature of coevolutionary rules molding the interactions among players is also the time scale separation between link and strategy adaptations, as reported in \citep{santosXploscb06, pachecoXjtb06, pachecoXprl06, szolnokiXepl08, pachecoXjtb08, vansegbroeckXbmceb08, fuXpre08, szolnokiXepl09, vansegbroeckXprl09, szolnokiXnjp09}. As the cited works suggest, the time scale separation can drastically influence the final output of coevolutionary games. This effect will also be discussed in the present review. In what follows, we will review the coevolutionary rules presented in some of these works more accurately.

In agreement with the actual time-line we start with the work of \citet{ebelXaxv02}, who proposed a coevolutionary rule in which a randomly chosen player $x$ is connected to a new neighbor at random. If the new link increases the average payoff of the focal player the latter accepts it and disconnects from the neighbor it scores worst against. Note that this coevolutionary rule indirectly favors the establishment of cooperator-cooperator links (this pair-up yields the highest average payoff) and at the same time facilitates the deletion of defector-defector links. The coevolutionary rule was paired up with strategy mutation \citep{ebelXpre02}, by which a mutation is accepted if it yields a higher payoff for player $x$ than the initial strategy [type B; see Fig.~\ref{fig:class_of_updates}(b)]. Starting from a random network with Poissonian degree distribution, it was shown that this coevolutionary rule leads to cooperative Nash equilibria in an iterative prisoner's dilemma game with the additional property that no agent can improve its payoff by changing its neighborhood. According to the authors, the later may be interpreted as a sort of `network Nash equilibrium' \citep{ebelXaxv02}. Notably, this coevolutionary rule also affects the initial network structure in that the later evolves to a statistically stationary state with a broad degree distribution, suggesting scale-free behavior and giving rise to small-world properties, among others.

Following their preceding seminal contribution [see \citet{zimmermannX01}], \citet{zimmermannXpre04} proposed a coevolutionary rule affecting only defector-defector pairs with the motivation that in this pair-up \textit{both} players might be better off if searching for a new partner in the context of the prisoner's dilemma game [type A; see Fig.~\ref{fig:class_of_updates}(a)].
It was shown that, starting from a random network with a given average degree and the richest-following strategy adoption rule, even a small probability $p$ of searching for a new partner from a defector-defector configuration may substantially promote cooperation. Indeed, as low as $p=0.01$ were shown to uphold practically complete cooperator dominance across the whole span of the weak prisoner's dilemma game (see Fig.~\ref{fig:schematic_games}). With respect to the network topology, it was reported that the coevolutionary rule facilitates the formation of a hierarchical interaction structure and may also introduce small-world properties if the search for new partners is constrained to the neighbors of the neighbors. However, unlike as shown by \citet{ebelXaxv02}, the occasional (depending on $p$) break-up of defector-defector pairs has not been found leading to broad or even scale-free degree distributions. These findings were subsequently extended \citep{zimmermannXpre05, eguiluzXajs05}, where it was elaborated on the spontaneous emergence of cooperators with extremely high payoffs and the important role of this so-called `leaders' for the global sustenance of cooperation. As such, these works can be considered as an important prelude to the realization of the fact that scale-free networks constitute an extremely favorable environment for the evolution of cooperation irrespective of the governing social dilemma \citep{santosXprl05, santosXpnas06}.

A simple but still plausible coevolutionary rule affecting links between players has been proposed by \citet{pachecoXjtb06, pachecoXprl06}.
Exemplifying type A class of interaction updating [see Fig.~\ref{fig:class_of_updates}(a)], players adopting either the strategy $C$ (cooperate) or $D$ (defect) were designated a propensity to form new links denoted by $\alpha_C$ and $\alpha_D$, such that $xy$ links were formed at rates $\alpha_x \alpha_y$, where $x,y \in [C,D]$. Moreover, each link was assigned a specific lifetime depending on the strategy of the two connected players given by $\tau_{xy}=\gamma_{xy}^{-1}$, where $\gamma_{xy}$ is the corresponding link death rate. With these definitions the authors were able to specify mean field equations governing the so-called active linking dynamics of the network. This coevolutionary rule has been tested on the prisoner's dilemma and the snowdrift game subject to the Fermi function (see Eq.~\ref{fermi}) governing the strategy adoption [for additional set-ups see \citet{pachecoXjtb06}]. It was shown that if the time scale associated with active linking is much smaller than the one associated with strategy updating the proposed coevolutionary rule leads to an effective rescaling of the governing payoff matrix, and thus a shift in the played evolutionary game. For example, the prisoner's dilemma game transforms to the coordination game, while the snowdrift game transforms to the harmony game [for details on the coordination game see \textit{e.g.} \citet{szaboXpr07}]. In both cases the cooperation is promoted, in turn designating the proposed coevolutionary rule as a simple and analytically tractable means of understanding how selfish and unrelated individuals may be led to adopting the cooperative strategy. On the other hand, if the ratio between the time scales associated with active linking and strategy updating is not small, the interplay between these two dynamical processes leads to a progressive crossover between the analytic results obtained for very fast active linking and the evolutionary dynamics of strategies taking place on static graphs. The latter were found to exhibit different degrees of heterogeneity depending on the parameters determining active linking, yet in general complying well with real social networks having fast decaying tails in their degree distributions. Notably, compared to the earlier works reviewed above, an important observation made in the two papers by \citet{pachecoXjtb06, pachecoXprl06} was that the impact of coevolutionary rules may depend significantly on the time scales associated with the strategy and structure (link) evolution. For example, \citet{zimmermannXpre04} too commented on the time scale separation in their model, yet the promotion of cooperation was thereby not notably affected [both slow ($p \ll 1$) and fast ($p \rightarrow 1$) rewiring of $D-D$ links was found to be highly effective]. Active linking dynamics has also been investigated in repeated games incorporating direct reciprocity \citep{pachecoXjtb08}, where additionally the productivity of every link connecting two players was evaluated prior to potential rewiring. Moreover, the active linking model proposed by \citet{pachecoXprl06} was recently extended by \citet{vansegbroeckXprl09} to account for the impact of different reactions to adverse ties. In particular, \citet{vansegbroeckXprl09} additionally introduced individual behavioral types of players through different values of $\gamma$, separating those that tend to break their links frequently ($\gamma$ close to $1$) from those that tend to break them rarely ($\gamma$ close to $0$). In this way both topology and strategy dynamics become interrelated. It was shown that populations in which individuals are allowed to handle their social contacts diversely are more prone to cooperative behavior than those in which such diversity is absent. Similarly as in \citet{pachecoXprl06}, it was shown that by an appropriate time scale separation between strategy and network dynamics the diverse behavioral preferences can also introduce a transformation of the governing social dilemma, yet so that each individual perceives the same game differently.

Also building on the time scale separation between rewiring and strategy updating is another paper by \citet{santosXploscb06}, where players are able to decide which links they want to maintain and which they want to change based on local information about their neighbors [type B; see Fig.~\ref{fig:class_of_updates}(b)]. A link change is initiated if player $x$ is dissatisfied with its connection to player $y$, which is the case if the strategy of player $y$ is to defect. However, player $y$ also assesses the quality of its link to $x$ in the same fashion. If both $x$ and $y$ are satisfied (which practically means that both are cooperators) the link between them remains intact. If $x$ wants to remove the link and $y$ not ($s_x=C$ and $s_y=D$), the probability $W$ given by the Fermi function (see Eq.~\ref{fermi}) is invoked. If realized, player $x$ is allowed to redirect to a random neighbor of $y$. If not, $x$ stays linked with $y$. If both $x$ and $y$ are defectors, and thus both want to remove the link, then rewiring takes place such that the new link keeps attached to $x$ with probability $W$ or to $y$ with probability $1-W$. Finally, the authors introduce a ratio defined as the time scale associated with the evolution of strategies $\tau_e$ (for simplicity equal to one) divided by the time scale associated with the rewiring of the network $\tau_a$, showing that there exists a critical value for this ratio above which cooperators wipe out defectors. Moreover, the emerging networks exhibit an overall heterogeneity that is maximal at the critical value and can be compared well with diversity associated with realistic social networks. The coevolutionary rule proposed by \citet{santosXploscb06} was extended by allowing individuals to adjust their social ties \citep{vansegbroeckXbmceb08}, with the extension that each player was assigned an individual willingness $0 \leq \eta \leq 1$ to rewire unwanted social interactions. Accordingly, players with small $\eta$ can be considered as loyal to their partners and resilient to change, while those with $\eta \rightarrow 1$ are swift in altering their links. It was shown that the highest cooperation levels can be achieved when the propensity to change links is highly strategy-dependent. More precisely, it was found very beneficial for the evolution of cooperation if defectors changed their partners frequently while cooperators behaved oppositely, \textit{i.e.} kept their partners for as long as possible. This is indeed expect since defectors are unable to establish social ties under mutual agreement with their partners. On the other hand, cooperators are typically much more prone to establishing long-term relations and loyalty. Ultimately, these two facts lead to the evolution of heterogeneous interactions networks where cooperators are known to prevail over defectors [see \textit{e.g.} \citet{santosXprl05}].

Related to the work of \citet{santosXploscb06} is the recent paper by \citet{fuXpre09}, the difference being that in the latter only cooperators are allowed to switch their partners if they act as defectors, and moreover, the new partner is sought randomly from the whole population. It is found that under such a coevolutionary rule there exists an optimal, rather than critical [compare with \citet{santosXploscb06}], time scale separation between rewiring and strategy updating for which cooperation thrives best. Also, the resulting interaction topology is different in that the network typically becomes divided into isolated communities of cooperators and defectors due to the selective coevolutionary rule targeting only mixed strategy pairs. Notably, a preceding study by \citet{fuXpre08} considered partner switching also with the aid of reputation, which was defined similar to image scoring proposed a decade earlier by \citet{nowakXn98}. It was found that coevolutionary switching of partners based on the reputation of nearest and next-nearest neighbors, \textit{i.e.} preferentially targeting players with a higher reputation, is significantly more effective in promoting cooperation than seeking a new partner randomly from the whole population. Thus, these results underline the importance of indirect reciprocity \citep{nowakXjtb98, nowakXn98} also when individuals can adjust their social ties.

\begin{figure}
\begin{center}
\includegraphics[width=7.5cm]{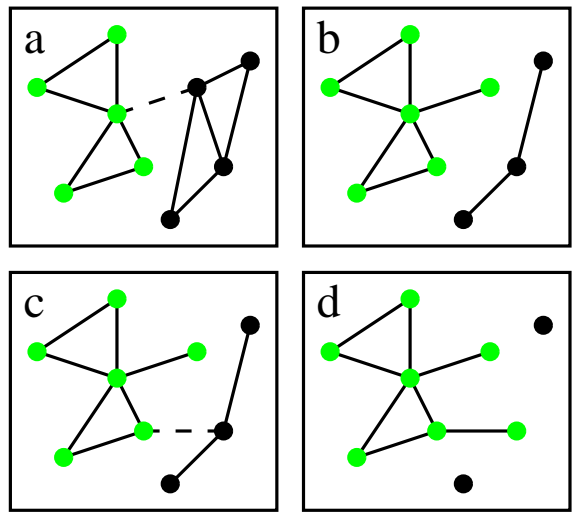}
\caption{Multilevel selection at work. In all panels cooperators (defectors) are denoted by green (black) circles. A cooperator, strengthened by neighboring cooperators (note that $C-C$ links are beneficial for all involved), can pass its strategy to a defector that is weakened by neighboring defectors (panel a). Subsequently, the invaded player looses its links to other players, except the one with the donor of the new strategy (panel b). Due to random link additions, the successful invasion of cooperators will repeat itself sooner or later depending on $\tau$ (panel c), ultimately resulting in the disintegration of the defector cluster (panel d). Note that this process cannot work in the opposite direction, \textit{i.e.} defectors cannot invade a cluster of cooperators. The necessary condition for this mechanism to work is the emergence of quasi-homogeneous groups, which occur if strategy adoptions happen frequently between new link additions, \textit{i.e.} if $\tau$ is large enough.}
\label{fig:multilevel_selection}
\end{center}
\end{figure}

In addition to the studies reviewed above, similar coevolutionary rules were used to study how scale-free networks emerge in social systems \citep{wliXpre07}, how cooperation in the prisoner's dilemma game can be established via the interplay between dynamical interactions and game dynamics \citep{fuXpa07} or interaction stochasticity \citep{xjchenXpre08b}, how social dilemmas in general can thereby be resolved \citep{tanimotoXpre07, pestelacciXbt08, tanimotoXpa09}, as well as other sophisticated models \citep{hanakiXms06, bielyXpd07} were considered. We refer the interested reader to the original works for further details, while here we proceed with the review of some of the studies that employed strategy independent rules for link adaptations \citep{szolnokiXepl08, percXpre08b, akunXbs09, tanimotoXieee09, szolnokiXepl09}.

Belonging to the third type of interaction-updating coevolutionary rules [type C; see Fig.~\ref{fig:class_of_updates}(c)] is the model proposed by \citet{szolnokiXepl09}, where whenever player $x$ adopts a new strategy all its links, except from the one with the donor of the new strategy, are deleted (see Fig.~\ref{fig:multilevel_selection}), and moreover, all individuals are allowed to form a new link with a randomly chosen player with which they are not yet connected after every $\tau$ full Monte Carlo steps. Note that the random additions of links counteract the deletions following each strategy adoption, in turn largely preserving the initially random topology and the heterogeneity of the interaction network \citep{szolnokiXnjp09}. It was shown that at a sufficiently large time scale separation between link deletions and additions, constituted by $\tau$, this coevolutionary rules evokes the spontaneous emergence of a powerful multilevel selection mechanism, which despite the persistent random topology of the evolving network, maintains cooperation across a substantial portion of the $T-S$ parameter plane. Importantly, the promotion of cooperation is thereby not realized by some final outcome of a coevolutionary rule, as is for example the case in \citep{szolnokiXepl08}, but is the consequence of a dynamical processes that affects the adoption of strategies on the macroscopic level of evolutionary game dynamics. As Fig.~\ref{fig:multilevel_selection} illustrates, the latter manifests as multilevel selection \citep{wilsonXbbs98, traulsenXpnas06} that strongly promotes cooperation in all major types of social dilemmas.

\begin{figure} [!ht]
\begin{center}
\includegraphics[width=7.0cm]{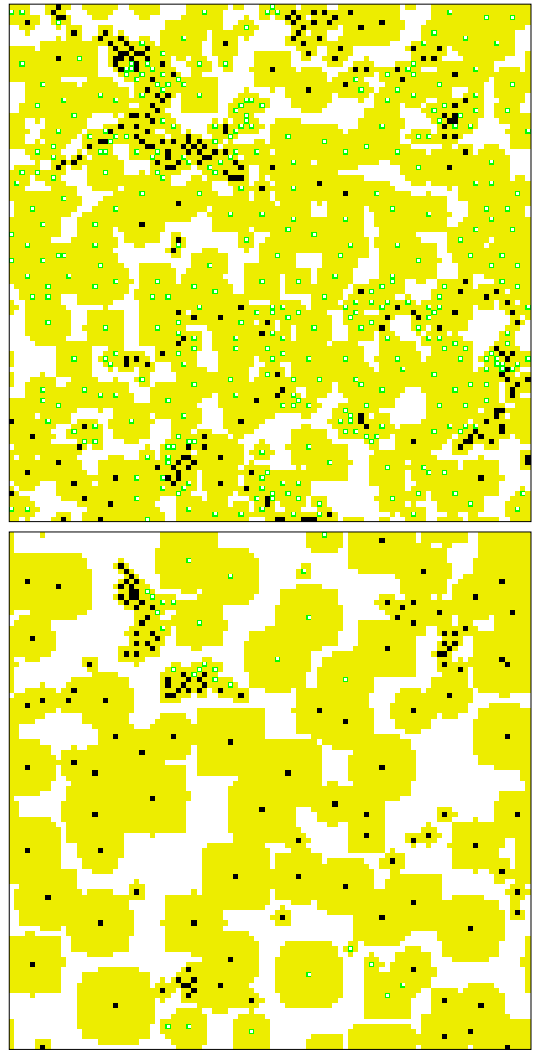}
\caption{Typical distributions of players on a $100 \times 100$ grid, obtained at an optimal ($k_{max} = 50$; top panel) and a too large ($k_{max} = 200$, bottom panel) connectivity originating from the coevolutionary rule proposed by \citet{szolnokiXepl08}. Full black (open green) boxes depict the positions of influential defectors (cooperators) while yellow (white) pixels depict the players who are within (out of) their range of influence. If the influential players are separated by large disjunct territories of influence (bottom panel) the network reciprocity is not functioning well.}
\label{fig:coev_range}
\end{center}
\end{figure}

Conceptually fitting to the third type of interaction-updating coevolutionary rules [type C; see Fig.~\ref{fig:class_of_updates}(c)] is also the model introduced by \citet{szolnokiXepl08}, where each player $x$ that successfully passes its strategy (\textit{i.e.} reproduces in a biological scenario) is allowed to form a new link with one randomly selected neighbor from its current neighborhood, thereby increasing its degree $k_x$ by one. Thus, successful players are allowed to grow compact large neighborhoods that are centered around their initial four nearest neighbors. As it is generally assumed, the payoff of any given player is accumulated from all the links with its neighbors. Hence, without the normalization by degree, the more links a player has the higher its payoff is expected to be. For the sake of an easier depiction of player distributions, we start from an interaction graph that can be represented by a square lattice. Evidently, the additions of new links will drive the initial topology away from two dimensions, yet still allowing us to capture relevant details of strategy distributions via a square lattice representation, as shown in Fig.~\ref{fig:coev_range}. Notably, the coevolutionary rule is independent on whether $s_x=C$ or $D$, and can hence be considered as strategy independent. However, since the performance of the strategies is clearly definitive for who gets to make new links, the rule has at least conceptual similarities with some of the above-reviewed works that considered strategies as more directly decisive for the outcome of dynamical interactions. Since the coevolutionary rule would eventually result in a fully connected graph (the latter constitutes well-mixed conditions), the parameter $k_{max}$ was introduced as the maximal degree a player is allowed to obtain. Accordingly, the process of making new connections is stopped as soon as the degree $k$ of a single player within the whole population reaches $k_{max}$, whereby this limit prevents the formation of a homogeneous system and indeed constitutes the main parameter affecting the impact of the coevolutionary rule. Starting from a square lattice, it was shown that intermediate values of $k_{max} \approx 50$ substantially promote cooperation in the weak prisoner's dilemma game (see Fig.~\ref{fig:schematic_games}) governed by Eq.~\ref{pach}, which was attributed to the formation of highly heterogeneous interactions networks ensuring optimal transfer of information between influential players, \textit{i.e.} those that have the highest degree among any other players that can adopt the strategy from the influential player via an elementary process. The coevolutionary rule was also tested against robustness to time scale separation between neighborhood growth and strategy adoption via the introduction of a parameter $q$, defining the probability of degree extension after a successful strategy pass. Evidently, $q=1$ recovers the originally proposed model while decreasing values of $q$ result in increasingly separated time scales. Although the impact of $q$ was found depending somewhat on the temptation to defect $b$, in general values of $q>0.2$ yielded insignificantly different results if compared to the $q=1$ case. Note that $q=0$ corresponds to the spatial model without coevolution, and hence it is natural that as $q \rightarrow 0$ the promotion of cooperation was found fading. The success of intermediate values of $k_{max}$ in promoting cooperation can be explained based on the emergence of heterogeneous interaction networks and the disassortative mixing of high-degree nodes \citep{rongXpre07, tanimotoXieee09}. In particular, while intermediate values of $k_{max}$ result in a highly degree-diverse mixture of players, which generally promotes cooperation [see also \citet{santosXprl05}], too large values of $k_{max}$ yield just a few influential players with disjunct clouds of homogeneous regions surrounding them, as shown in the bottom panel of Fig.~\ref{fig:coev_range}. In the later case, the lack of information exchange between hubs (influential players having large degree) defectors can easily survive, thus resulting only in moderate cooperation levels. The top panel of Fig.~\ref{fig:coev_range}, on the other hand, features an optimal distribution of influential players (\textit{i.e.} those having large degree), where high-degree cooperators can make cooperation prevail practically across the whole system.

\begin{figure}
\begin{center}
\includegraphics[width=7.5cm]{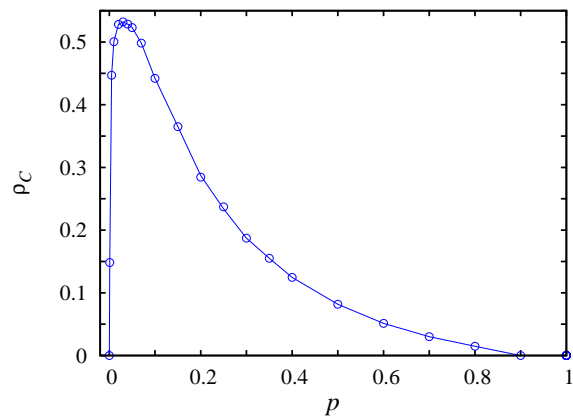}
\caption{Fraction of cooperators $\rho_C$ as a function of $p$ in the `connected influential players' model \citep{percXpre08b}. Parameter $p$ determines the intensity of the information exchange between influential players. The fraction of influential players is $\mu = 0.12$ and the temptation to defect is $b = 2$. The line is just to guide the eye.}
\label{fig:restricted}
\end{center}
\end{figure}

It is worth mentioning that the optimal level of cooperation observed for an intermediate value of $k_{max}$ in \citet{szolnokiXepl08} is conceptually similar to the case when an intermediate strength of information exchange between influential players yields the optimal environment for cooperation \citep{percXpre08b}, as is illustrated in Fig.~\ref{fig:restricted}. In the later case a fraction $\mu$ of players that are characterized with a larger teaching capability are allowed to temporarily link with distant opponents of the same kind with probability $p$, thus introducing shortcut connections among the distinguished. These additional temporary connections are able to sustain cooperation throughout the whole range of the temptation to defect $b$ (see Fig.~\ref{fig:schematic_games}). As Fig.~\ref{fig:restricted} demonstrates, only minute values of $p$, constituting a moderate intensity of information exchange between influential players, warrant the best promotion of cooperation.

Strategy-independent coevolutionary rules affecting the interactions between players have also been considered in the context of distinguished players populating a square lattice \citep{percXpre08b}, in the context of dynamically changing random and scale-free networks \citep{akunXbs09}, as well as in the context of random networks with different assortative mixing emerging due to links adaptations \citep{tanimotoXieee09}. Again, the interested reader is referred to the original works for further details, while here we proceed with a new section devoted to the review of coevolutionary rules introducing network growth.

\subsection{Population growth}
\label{growth}

First, it is worth noting that coevolutionary rules giving rise to population growth have been considered much less frequently than the above-reviewed rules affecting solely how players link with one another. Indeed, only three works fit into this subsection, the latter being the arXiv contribution by \citet{renXcm06b} and two recent papers by \citet{poncelaXploso08, poncelaXnjp09}. Closely related to the rather general and broad interest in network growth \citep{grossXjrsi08, castellanoXrmp09}, the networks formed by the players participating in evolutionary games can be subject to growth as well, with motivations equivalent to those of the broader research field.

Although never officially published, the work by \citet{renXcm06b} should be acknowledged as being pioneering in raising the question how the dynamics of an evolutionary game might affect network growth, and how in turn the latter affects back the prevalence of the competing strategies. For this purpose, the authors proposed a so-called payoff-based preferential attachment rule under the guidance of the $r$-parameterized snowdrift game (see Fig.~\ref{fig:schematic_games}) and the Fermi strategy adoption rule given by Eq.~\ref{fermi}. Indeed, the preferential attachment rule by \citet{renXcm06b} is practically identical to the seminal growth and preferential attachment model proposed by \citet{barabasiXs99}, only that in the former the probability of linking a new player to an existing player $x$ is not determined by its degree $k_x$ but rather by its accumulated payoff until that time. Not surprisingly then (note that in the absence of normalization, similarly as the degree of a player, its payoff will typically also increase by one during an update), the coevolutionary rule was found leading to the emergence of scale-free interaction networks that are characterized by the degree distribution $P(k) \propto k^{- \gamma}$; the coefficient $\gamma$ thereby depending on the scaling of the probability of linking a new player to an existing player. In accordance with an earlier study by \citet{santosXprl05}, the emerging scale-free topology due to the coevolutionary rule was found highly beneficial for the evolution of cooperation in the snowdrift game. Notably, the authors also investigated the average path length and the assortative mixing of the emerging networks, as well as the wealth distribution of players. The former two were found to be in agreement with observations from realistic social networks, while the latter was found consistent with the Pareto law.

The work by \citet{poncelaXploso08} also introduces an evolutionary preferential attachment rule that is based on the payoffs of existing players, albeit the weak prisoner's dilemma is employed as the governing game and the strategy adoption probability is quantified according to Eq.~\ref{pach}. More precisely, the network growth starts with $m_0=3$ fully connected players and proceeds by adding a new player with $m=2$ links to the existing ones at equally spaced time intervals $\tau_T$. The probability that any player $x$, (having payoff $\Pi_x$) in the network receives one of the $m$ new links was defined as
\begin{equation}
p_x(t)= \frac{1-\epsilon+\epsilon \Pi_x(t)}{\sum_y [1-\epsilon+\epsilon \Pi_y(t)]}
\label{attach}
\end{equation}
where the sum runs over all the players forming the network at time $t$. Moreover, the parameter $\epsilon \in [0,1)$ controls the weight of the payoffs during the network growth. For $\epsilon=0$ all nodes are equiprobable, corresponding to the weak selection limit [see \textit{e.g.} \citet{traulsenXjtb07b, wildXjtb07, fuXpre09b} for recent works related to the latter], while for $\epsilon \rightarrow 1$ the players with the highest payoffs are much more likely to attract the newcomers. The authors also specified the time interval $\tau_D$ for payoff evaluations and potential strategy adoptions, focusing explicitly on $\tau_D / \tau_T >1$ (typically $\approx 10$, although smaller and larger values were also commented on), so that accordingly the network growth was considered to be faster than the evolutionary dynamics. It was shown that the weak selection limit results in networks having degree distributions with exponentially decaying tails, while the strong selection limit ($\epsilon \rightarrow 1$) yields highly heterogeneous scale-free interaction networks. In agreement with the earlier findings obtained on static graphs \citep{szaboXpr07}, it was confirmed that higher levels of cooperation are attainable on heterogeneous rather than homogeneous topologies, albeit that the distribution of strategies with respect to the degree of nodes forming the network is different. More precisely, cooperators were not found occupying the main hubs as on static graphs, but rather the nodes with an intermediate degree, thus indicating that the interplay between the local structure of the network and the hierarchical organization of cooperation is guided by the competition between the network growth and the evolutionary dynamics. Notably, similar differences in the microscopic organization of the steady state composition of strategies were found on static scale-free networks when the payoffs were subjected to normalization \citep{szolnokiXpa08}, although the discrepancies reported by \citet{poncelaXploso08} were solely the consequence of the coevolutionary growth process. Indeed, in a recent study \citet{poncelaXnjp09} this coevolutionary rule has been studied further to confirm that the reported promotion of cooperation hinges not only on the final heterogeneity of the resulting network but also vitally on the particularities of the growth process itself. In addition, it was shown that under strongly payoff dominated growth conditions so-called super-hubs can emerge, which attract most of the links from the other nodes. Although under such conditions cooperation was found thriving even for high temptations to defect, it was also noted that the robustness of these findings may be compromised, or at least not so strong as on static scale-free networks \citep{poncelaXnjp07}, due to the extreme heterogeneity of the star-like structures that can be brought about by the coevolutionary network growth.

With the above we conclude the review of coevolutionary rules affecting the interaction network, either in terms of links players form with one another (see Section~\ref{interactions} above) or the actual number of players participating in the game and the related network size. We proceed with the review of coevolutionary rules affecting individual properties of players, such as their teaching activity (see Section~\ref{teaching}), mobility (see Section~\ref{mobility}) or age (see Section~\ref{aging}). Note, however, that some of the above-reviewed coevolutionary rules already incorporated and/or affected personal features of players, such as for example the loyalty to their partners \citep{vansegbroeckXbmceb08, vansegbroeckXprl09} or influence \citep{percXpre08b}, albeit always in conjunction with the coevolution of the interaction network. In what follows, the links and the size of the network are not affected by the coevolutionary rules unless explicitly noted otherwise.

\subsection{Evolving teaching activity}
\label{teaching}

Heterogeneity of players has been explored as a beneficial condition for cooperation in several forms \citep{zxwuXcpl06, percXpre08, fortXpa08, masudaXjtb08}. It can be easily accepted that players are not perfectly identical within a population. Some have higher reputation or stronger influence than others. These differences can be detected via a biased direction of strategy adoptions. More precisely, players with higher reputation can spread their strategy more easily than if having an average or low reputation. In other words, their activity to teach a neighbor a new strategy is higher. It turned out that one of the individual quantities that influences the evolution of cooperation most effectively is the teaching activity \citep{szolnokiXepl07}. Notably, teaching activity can also be referred to as the influence or reproduction rate \citep{szolnokiXepjb08}, with the logical assumption that influential individuals are much more likely to reproduce, \textit{i.e.} have a higher teaching activity, than players with low influence. Teaching activity (or the synonyms we pointed out) can be introduced into the framework of evolutionary game theory via a modified Fermi strategy adoption rule
\begin{equation}
W(s_x \rightarrow s_y)= w_x \frac{1}{1+\exp[(\Pi_y-\Pi_x)/K]}
\label{teachfermi}
\end{equation}
where $w_x$ characterizes the strength of influence (or teaching activity) of player $x$. Obviously, $w_x=1$ for all $x$ returns Eq.~\ref{fermi}, whereby it is important to acknowledge that even if $w_x<1$ but the same for all $x$ the evolutionary outcome of strategy abundance remains the same, only the relaxation times lengthen. Quenched (non-evolving) distributions of $w_x$ may promote cooperation even on homogenous lattice-type interaction topologies \citep{szolnokiXepl07}, while their application on complex networks reveals further that players with large teaching activity play a similar role as hubs in highly degree heterogenous graphs, such as scale-free networks \citep{szolnokiXepjb08}. We refer the reader to the original works for further details on models using quenched distributions of $w_x$, while here we proceed with the review of the two papers by \citet{szolnokiXnjp08, szolnokiXepjb09} that thus far considered the teaching activity as an evolving property of individual players.

\begin{figure}
\begin{center}
\includegraphics[width=7.5cm]{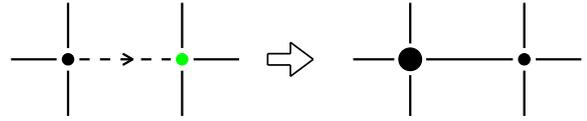}
\caption{Coevolution of teaching activity during a strategy adoption, as proposed by \citet{szolnokiXnjp08}. The teaching activity of the left player, which is proportional with the size of circle, increases due to the successful strategy pass. Note that the right player adopts the strategy from the left player, hence the change of color from green to black. This strategy-independent (note that the teaching activity of the left player increases irrespective of which strategy was passed to the right player) coevolutionary rule can result in highly heterogenous distributions of teaching activity (see Fig.~\ref{fig:teaching_dist}), which were found beneficial for the evolution of cooperation irrespective of the underlying interaction network.}
\label{fig:rule_evol_teach}
\end{center}
\end{figure}

In a social context the strategy adoption can be considered as learning from the more successful player. Accepting this point of view, it is straightforward to consider a player who has successfully passed a strategy as the one having a higher reputation, and thus a higher teaching activity than other players. Implementing this idea into a coevolutionary rule, we proposed that whenever player $x$ successfully passes its strategy the influence $w_x$ increases by a constant positive value $\Delta w \ll 1$ according to $w_x \rightarrow w_x + \Delta w$ \citep{szolnokiXnjp08}. This coevolutionary rule is illustrated in Fig.~\ref{fig:rule_evol_teach}. It should be noted that in this model the term `reputation' does not necessarily have a positive meaning, and thus may be in contradiction with the same term used elsewhere \citep{fuXpre08}, where players who cooperated were awarded a higher reputation, which expectedly yielded higher levels of cooperation.

\begin{figure}
\begin{center}
\includegraphics[width=7.5cm]{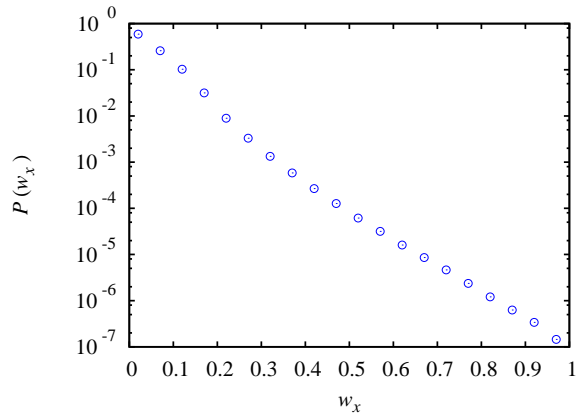}
\caption{Spontaneously emerging heterogeneous distribution of the teaching activity $P(w_x)$ as a result of the coevolutionary rule introduced by \citet{szolnokiXnjp08}. Results were obtained for the weak prisoner's dilemma game staged on a square lattice. Parameter values were: $b=1.05$, $K=0.1$ and $\Delta w = 0.07$.}
\label{fig:teaching_dist}
\end{center}
\end{figure}

Moreover, for the sake of simplicity it was assumed that the evolution of $w$ stops as soon as the highest $w_x$ reaches $1$ \citep{szolnokiXnjp08}. Starting from a nonpreferential setup, initially assigning $w_x=0.01$ to every player irrespective of its strategy, it was found that there exists an optimal intermediate value of $\Delta w \approx 0.07$ for which cooperation in the weak prisoner's dilemma as well as the $r$-parameterized snowdrift game (see Fig.~\ref{fig:schematic_games}) is enhanced best. It is in fact understandable that only an intermediate value of the increment $\Delta w$ was found warranting the optimal heterogeneity of the distribution of $w_x$. Namely, if $\Delta w$ is small then the values of $w_x$ simply increase homogeneously for all the players, while large values of $\Delta w$ result in a very quick halt of the coevolutionary process, either way resulting in a rather homogeneous distribution of the teaching activity. Indeed, for both considered evolutionary games it was found that using moderate $\Delta w$ the final distribution of $w$ is exponential, in turn attributing the promotion of cooperation to the spontaneously emerging highly heterogenous plethora of differently influential players, as shown in Fig.~\ref{fig:teaching_dist}. It was also shown that the effectiveness of the coevolutionary rule increases with the increasing uncertainty by strategy adoptions $K$, and that the rule is robust to variations of the updating scheme. For example, it was verified that an alternative coevolutionary rule, by which $w_x$ was allowed to grow also past $1$ only that then $w_x$ was normalized according to $w_x \rightarrow \frac{w_x}{w_{max}}$ ($w_{max}>1$ being the maximal out of all $w_x$ at any given time) to ensure that the teaching activity remained bounded to the unit interval, yielded similar results as the halted version.

A two-fold extension of the above work was made in \citet{szolnokiXepjb09}.
First, the coevolutionary rule was no longer considered to be strategy independent. Note that in the preceding work $w_x \rightarrow w_x + \Delta w$ was executed irrespective of the strategy of player $x$. Conversely, in \citet{szolnokiXepjb09} this rule was applied separately either only for $s_x=C$ (cooperators) or only for $s_x=D$ (defectors). Second, the evolution of cooperation was examined in all three major social dilemma types defined on the $T-S$ parameter plane (see Fig.~\ref{fig:schematic_games}). It was shown that both versions of the coevolutionary rule promote cooperation irrespective of the underlying game. Opposite to intuitive reasoning, however, it was revealed that the exclusive coevolutionary promotion of players spreading defection is more beneficial for cooperation than the likewise direct promotion of cooperators. This was attributed to the fact that the coevolutionary promotion of defectors results in a larger fraction of players that are at least once affected by the coevolution, ultimately leading to a stronger segregation of the population into active (those having $w_x > 0.01$; note that the latter is the initial teaching activity assigned to all) and virtually (or comparably) inactive (those having $w_x = 0.01$) players than the coevolutionary rule affecting cooperators. According to previous findings on the impact of static distributions of heterogeneity \citep{percXpre08}, the stronger expressed segregation was found directly responsible for the better promotion of cooperation when defectors rather than cooperators were subjected to coevolution.

As we have already mentioned (see \textit{e.g.} Section~\ref{interactions}), the time scale separation of coevolutionary processes may decisively affect the final output of such models. This was observed for the coevolution of teaching activity as well. More precisely, the time scale separation between the coevolution of teaching activity and strategy adoption can be tuned via the introduction of a parameter $q$, defining the probability of increasing $w_x$ after a successful strategy pass. Evidently, $q=1$ recovers the two originally proposed models while decreasing values of $q$ result in increasingly separated time scales. Although the impact of $q$ was found depending somewhat on the type of the considered coevolutionary rule, in general, values of $q>0.3$ yielded insignificantly different results if compared to the $q=1$ case, thus indicating that the findings are robust to this type of alterations [note that, as in \citet{szolnokiXepl08}, $q=0$ corresponds to the model without coevolution, and hence it is natural that as $q \rightarrow 0$ the promotion of cooperation was found fading]. More precisely, however, since the fraction of cooperators was found increasing rather steadily with increasing values of $q$, especially for the coevolutionary rule affecting defectors, it is optimal to keep the coevolutionary process affecting the teaching activity of players paced similarly fast as the main evolution of strategies, \textit{i.e.} $q \rightarrow 1$.

We thus emphasize, that the above-reviewed coevolutionary models affecting the teaching activity have revealed that a simple `successful become more successful' principle can result in a heterogenous hierarchy of individual properties of players, such that optimal conditions for the evolution of cooperation are warranted. A similarly positive impact of heterogeneity on the spread of the cooperative strategy was also detected on heterogenous interaction networks \citep{santosXprl05}, hence conceptually linking these two seemingly disjoint promoters of cooperation.

\subsection{Mobility of players}
\label{mobility}

It was acknowledged already by \citet{majeskiXc99} that players finding themselves in an unprofitable or undesirable situation frequently choose moving in order to free themselves from the negative consequences of that situation. Accordingly, mobility can be considered as being a coevolutionary process in the sense of strategy and/or position alterations that ultimately determine the environment of players. Although we were unable to locate coevolutionary terminology associated with mobility, we review here advances on this topic made during the last decade, and indeed consider movements of players during the evolution of strategies as being guided by rules of coevolution.

The impact of diffusion on the outcome of a spatial prisoner's dilemma game via empty sites was questioned first by \citet{vainsteinXpre01}. Therein, weak quenched disorder introduced in the form of empty sites on a square lattice was found beneficial for cooperation in the prisoner's dilemma game subject to the richest-following strategy update rule. In a follow-up paper \citep{vainsteinXjtb07} the approach was extended to allow diffusion of players to nearest-neighbor empty sites with a certain probability. In particular, two ways of implementing the mobility were considered. First, each player was allowed to make an attempt at moving only after payoff accumulation and potential strategy adaptations were executed in parallel, or second, the moving was attempted prior to the evolution of strategies. Importantly, the moving of players was considered to be Brownian random walk like, \textit{i.e.} diffusive, not relying on any type of explicit, genotypic or phenotypic assortment, and also being strategy-independent. Due to this minimalist set-up the study provided rather general insights into possible effects of mobility. It was shown that mobility may indeed promote cooperation since it increases the ability of cooperator clusters to invade and overtake isolated defectors. On the other hand, mobility may also allow defectors to escape retaliation from a former partner and lead to stronger mixing in a population due to increasing interaction ranges of players, both of which are known to damp the evolutionary success of cooperators. Thus, the impact of mobility in the form introduced by \citet{vainsteinXjtb07} is not clear cut. As noted by the authors, further work on this is in progress. Importantly, it was also emphasized that mobility may be subject to more deliberate coevolutionary rules, taking into account personal preferences of players, their strategies, as well as aims.

An example of the latter was studied by \citet{helbingXpnas09}, who introduced success-driven migration as a possible mediator leading towards cooperation in populations of selfish and unrelated individuals even under noisy conditions. In particular, success-driven migration [see also \citet{helbingXacs08}] was implemented so that, before the strategy adoption, player $x$ was allowed to explore potential payoffs that it would receive if occupying one of the empty sites in the migration neighborhood. The latter was typically restrained to nearest and next-nearest neighbors of player $x$. If the potential payoff was found to be higher than in the current location, player $x$ moved to the site offering the highest payoff and, in case of several sites with the same payoff, to the closest one. On the other hand, if the current location offered the highest payoff among all the empty sites within the migration neighborhood, player $x$ did not move. It was found that this fairly simple and very plausible migration rule promotes cooperation in the prisoner's dilemma game on a square lattice (with a fraction of empty sites to accommodate moving) irrespective of the noise introduced to the system. In fact, three types of noise were considered to attest to the robustness of cooperation facilitation due to the introduced mobility of players. First was the introduction of mutation with probability $q$, second was the introduction of random movements not considering the expected success (payoff) with probability $r$, while third was the combination of the two. Additionally, different update rules, adding birth and death processes, as well as introducing a small fraction of individuals defecting unconditionally were considered as well. Irrespective of all these factors cooperation was found prevailing in a large region of the parameter space defining the prisoner's dilemma game if only the players were allowed to execute success-driven migration.

Recently, mobility is getting increasing attention as a means to promote cooperation in social dilemmas \citep{drozXepjb09, meloniXpre09}. In the model of \citet{drozXepjb09} two types of players are introduced, and a random walk of the influential individuals is possible irrespective of their strategies. The mobility of influential cooperators can have two positive impacts on the evolution of cooperation. First, they can spread the cooperative strategy among the non-influential players having a lower teaching activity, and second, when two influential players with opposite strategies meet, the cooperator can prevail and thus ensure an effective information exchange between cooperating hubs. Note that the importance of the latter has as already been emphasized in Section~\ref{interactions} (see \textit{e.g.} Fig.~\ref{fig:restricted} and the pertaining text). As expected based on preceding works considering mobility as a coevolutionary process, the final outcome of the competition between mobile, and thus influential, cooperators and defectors is highly sensitive to changes in the speed of moving. When the latter is too high the influential cooperators cannot benefit from their cooperative neighborhoods because they abandon them too soon. Similarly, influential defectors eschew the negative feedback effect originating from defecting neighbors (note that $D-D$ links are nonprofitable for both players) because they leave them too fast as well. Indeed, high moving speeds generate conditions mimicking the well-mixed regime which is damning for cooperators. Thus, only moderate mobility of influential players has been found effectively supporting the evolution of cooperation. We refer the reader to the original works for further details, noting that mobility seems a promising avenue of research for future explorations of coevolutionary rules.

\subsection{Aging of players}
\label{aging}

As the last coevolutionary process we consider aging. Indeed, aging is always present, tailoring our interactions with others and postulating a finite lifespan during which we are able to exercise them. It thus seems natural to consider aging as an integral part of every evolutionary development, and certainly evolutionary games constitute a prime example thereof. Nonetheless, studies taking aging explicitly into account within this realm of research are few. \citet{mcnamaraXn08} recently noted that lifespan might play an important role in the evolution of cooperation, albeit their study focused on the coevolution of choosiness (see also Section~\ref{related}) rather than age. Moreover, concepts similar to aging were recently introduced in the voter model \citep{starkXprl08, starkXacs08}, showing that age and memory-dependent transition rates can have a positive effect on consensus formation. A recent work focusing on aging within evolutionary games is due to \citet{szolnokiXpre09}, and in the following we present a summary of the proposed coevolutionary rules for aging as well as their main implications.

\begin{figure}
\begin{center}
\includegraphics[width=7.5cm]{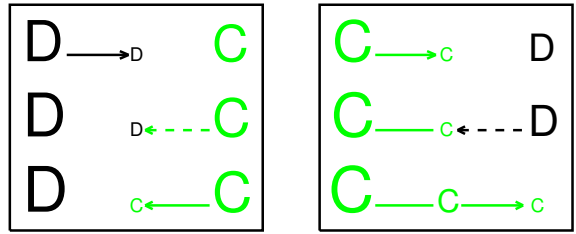}
\caption{Dynamical explanation of cooperation promotion emerging due to aging, as proposed by \citet{szolnokiXpre09}. When an old defector, with a high strategy transfer capability, is imitated by one of the neighbors, further spreading of defection is blocked because the newborn defector has no chance to pass strategy $D$ further. The newborn is not supported by its ancestor (the $D-D$ link is detrimental for both), and hence a neighboring cooperator with high age can conquer the site of the newborn defector. This procedure occurs repeatedly, ultimately resulting in a practically blocked (more precisely an oscillating) front between $C$ and $D$ regions (left panel). Importantly, a similar blocking is not present around old (and thus influential) cooperators because their cooperator-cooperator links help newborn cooperators to achieve higher age, in turn supporting the overall maintenance of cooperative behavior (right panel). In both panels the size of players is proportional to their age (\textit{i.e.} their teaching activity). Dashed (solid) arrows denote attempted (successful) strategy adoption processes.}
\label{fig:blocked}
\end{center}
\end{figure}

\begin{figure}
\begin{center}
\includegraphics[width=7cm]{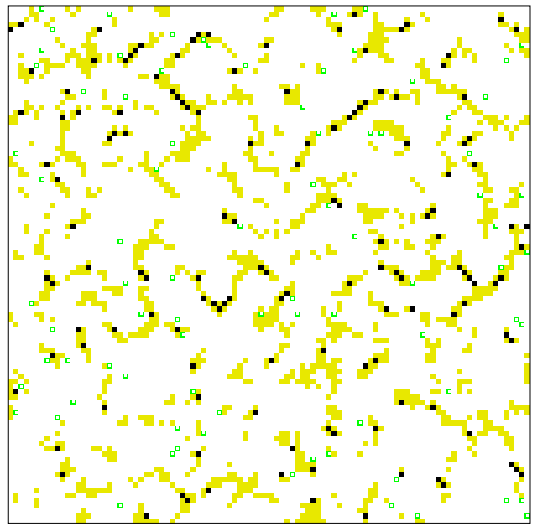}
\caption{Snapshot of a typical distribution of players on a $100 \times 100$ square lattice, obtained by considering players who have adopted a new strategy as newborns [coevolutionary rule B in \citet{szolnokiXpre09}]. Full black (open green) boxes depict influential defectors (cooperators) while yellow and white are all other non-influential defectors and cooperators, respectively. The snapshot demonstrates clearly that the propagation of defectors is blocked in space as a consequence of age-related teaching activity.}
\label{fig:aging_dist}
\end{center}
\end{figure}

Since age is often associated with knowledge and wisdom an individual is able to accumulate over the years, it was introduced through a simple tunable function that maps age to teaching activity (see Section~\ref{teaching}) of the corresponding player. More precisely, $w_x$ in Eq.~\ref{teachfermi} was related to the integer age $e_x=0, 1, \ldots, e_{max}$ in accordance with the function $w_x=(e_x/e_{max})^\alpha$, where $e_{max}=99$, denoting the maximal possible age of a player, serves the bounding of $w_x$ to the unit interval, and $\alpha$ determines the level of heterogeneity in the $e_x \rightarrow w_x$ mapping. Evidently, $\alpha=0$ corresponds to the classical (homogeneous) spatial model with $w_x=1$ characterizing all players, $\alpha=1$ ensures that $w_x$ and $e_x$ have the same distribution, whereas values of $\alpha \geq 2$ impose a power law distribution of strategy transfer capability. Although different age distributions of players were considered also as quenched system states, the focus was on the study of aging as a coevolutionary process, entailing death and newborns. Two rules were considered separately, both starting with $e_x$ being assigned randomly from a uniform distribution within the interval $[0,e_{max}]$ to all players, and subsequently increasing all $e_x$ by $1$ after each full Monte Carlo step. Also inherent to both was that $e_x$ was set to zero for all players whose age exceeded $e_{max}$ (effectively this means that a newborn follows the dead player). The difference was in the way age of players that have just adopted a new strategy from one of their neighbors was handled. In the coevolutionary model A their age was left unchanged, while in the coevolutionary model B they were considered as newborns, \textit{i.e.} as soon as player $x$ adopted a new strategy its age was set to $e_x=0$. Notably, rules A and B can be interpreted rather differently. From a purely biological viewpoint the more successful player replaces the neighbor with its own offspring, who therefore initially has a limited strategy transfer capability, which corresponds to rule B. On the other hand, especially in social systems, strategy adoptions may not necessarily involve death and newborns, but may indicate solely a change of heart, preference, or way of thinking, whereby this situation corresponds to rule A. Nevertheless, newborns in a social context can be considered those that changed their strategy recently, and therefore have a low reputation initially. Interestingly, it was found that the small difference between coevolutionary rules A and B may have significant consequences for the evolution of cooperation. Foremost, it was found that rule B promotes cooperation remarkably better than rule A. However, the difference could not be explained by the resulting heterogeneity of the distributions of $w_x$, for example via a similar reasoning as introduced by \citet{percXpre08}, since both rules return power law distributed values with rather similar slopes ($-0.5$ for A and $-0.7$ for B). In fact, it was shown that the coevolutionary rule B introduces a new powerful mechanism for promotion of cooperation acting solely on a microscopic player-to-player basis, and as such is thus virtually not detectable by statistical methods assessing the heterogeneity of the system. The mechanism was found relying on a highly selective promotion of cooperator-cooperator and defector-defector pairs, which hinders influential defectors (those having $e_x$ close to $e_{max}$) to spread their strategy effectively across the spatial grid. In particular, rule B always leads to influential players being surrounded by newborns. Thereby it is important to note that whenever an old defector, with a high strategy transfer capability $w_x$, is imitated by one of the neighbors, further spreading of defection is blocked because the newborn defector has no chance to pass strategy $D$ further. At that time a neighboring cooperator with high age can strike back and conquer the site of the newborn defector. As a result the whole procedure starts again, which ultimately results in a practically blocked (more precisely an oscillating) front between $C$ and $D$ regions. Crucially, a similar blocking mechanism is not present around old (and thus influential) cooperators because their cooperator-cooperator links help newborn cooperators to achieve higher age, in turn supporting the overall maintenance of cooperative behavior. The main differences in the propagation of different strategy pair-ups are summarized in Fig.~\ref{fig:blocked}, while an example of the resulting spatial distribution of players is presented in Fig.~\ref{fig:aging_dist}. In the latter a player is considered as influential if its age exceeds that of any of its neighbors by at least $e_{max}/2$ (qualitatively similar snapshots can be obtained by choosing different thresholds as well).

In model A the situation is significantly different since cooperative domains, created around old players with high $w_x$, cannot prevail long. Namely, the central cooperator who built up the cooperative domain eventually dies, and the arriving newborn with an accordingly low strategy transfer capability simply cannot maintain this domain further, thus giving defectors an opportunity to win it over. As a consequence of the dynamical origin of the observed cooperation-promoting mechanism brought about by rule B, it is expected that it will work in other cases too, for example when the interaction graph is characterized by a different topology, by other evolutionary games, or by separated time scales between aging and strategy adaptations, as was shown already by \citet{szolnokiXpre09}.

Finally, it is important to note that the observed cooperation-promoting mechanism relying on a dynamical process is robust even if non-monotonous mappings between $e_x$ and $w_x$ are considered (the oldest individuals may not necessarily be the most influential). Indeed, the promotion of cooperation remains intact as long as the plausible assumption that very young players should have none or very little influence is adhered to.

Similarly as mobility reviewed in Section~\ref{mobility}, we note that aging as a coevolutionary process seems very liable to further studies as well, and we hope this brief summary succeeded in wetting the appetite for them.

\subsection{Related approaches}
\label{related}

Aside from thus far reviewed coevolutionary rules, there exist examples \citep{kirchkampXjebo99, gintisXjtb03, axelrodXe04, hamiltonimXprsb05, fortXepl08, hatzopoulosXpre08, dingXijmpc09, moyanoXjtb09, scheuringXjtb09, rankinXev09, szaboXepl09} we were unable to classify into the above subsections. Without going into much details as it exceeds the scope of this mini review, we briefly describe some of these related approaches, but refer the reader to the original works for further details.

\citet{kirchkampXjebo99}, for example, studied the simultaneous evolution of learning rules and strategies, whereby the former were determined endogenously based on the success of strategies observed in the neighborhood of any given player. It was shown that endogenous learning rules put more weight on the proper understanding of each player's own experience rather than on the experience of an observed neighbor. Coevolving learning rules were recently considered also by \citet{moyanoXjtb09}, showing that imitation is frequently displaced by replication, in turn leading to a rapid decrease of cooperation in the spatial prisoner's dilemma game. On the other hand, imitation was found to be superior to global but stochastic imitation, thereby facilitating cooperative behavior. The coevolutionary selection of strategy adoption rules was consider by \citet{szaboXepl09} as well, where the uncertainty in the Fermi function (see Eq.~\ref{fermi}) was subject to evolution as a player-specific property. In particular, instead of a single $K$ value authors introduced different $K_i$ values where $i \in (1,2 \ldots, n)$, which were then assigned randomly to the players. As we have already noted in Section~\ref{games}, the uncertainty by strategy adoptions can originate from different sources, ranging from unpredictable variations in payoffs to errors in the decision making \citep{vukovXpre06, percXnjp06a, wbduXcpl09, zxwuXpre09, wbduXpa09}. The parameter $K$, however, can also be considered as characterizing the willingness of a player to risk a payoff quantity during a strategy change. Therefore, by using different values of $K$, not only the better strategy but also the way of strategy adoption can be the subject of an imitation process.

\begin{figure}
\begin{center}
\includegraphics[width=7.5cm]{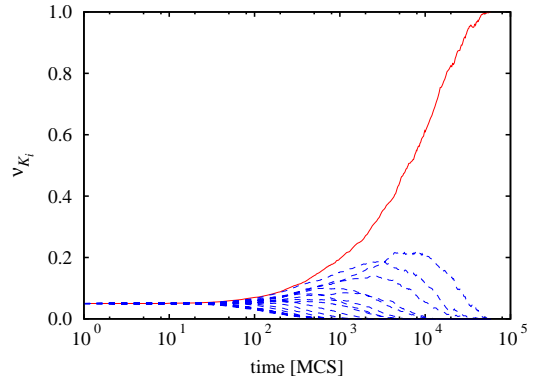}
\caption{Time dependencies of the fraction of initial $K_i$ values ($\nu_{K_{i}}$) demonstrate the selection of the most appropriate strategy adoption uncertainty $K^{\star}$ (promoting cooperation best), as indicated by the red solid line. This directly implies the spontaneous selection of an optimal strategy adoption rule within the scope of the Fermi function (see Eq.~\ref{fermi}), as reported by \citet{szaboXepl09}. Dashed blue lines depicted the extinction of the other $K_i$ values. Initially, $n=20$ different $K_i$ values were distributed on the square lattice of size $N=1000^2$. The temptation to defect was $b=1.05$.}
\label{fig:n20_time}
\end{center}
\end{figure}

Accordingly, aside from the fact that player $y$ could adopt the strategy of player $x$ according to Eq.~\ref{fermi} ($s_x \rightarrow s_y$), an additional independent trail with the same probability was made also for the adoption of the imitation rule ($K_x \rightarrow K_y$). It was shown that, if the system is seeded by random initial conditions, the proposed coevolutionary rule drives the system towards a state where a single $K^{\star}$ value (only one of the initial $K_i$ values) prevails. This final strategy adoption uncertainty is closely related with the parameter value warranting the highest cooperation level if a given value of $K$ is used for all players at a certain value of the temptation to defect $b$ (see Fig.~\ref{fig:schematic_games}). Naturally, the prevailing $K^{\star}$ value thus depends also on the topology of the interaction network. The selection process is illustrated in Fig.~\ref{fig:n20_time}, where $n=20$ different $K_i$ values were initially assigned to the players $x \in (1,2 \ldots, N)$. Summarizing the main observation, it was shown that a Darwinian selection rule affecting a model parameter can spontaneously lead to the prevalence of the value that ensures
an optimal level of cooperation in the system. For further details we refer the reader to the original work of \citet{szaboXepl09}.

The evolution of altruistic behavior under coevolutionary rules was studied also in what can be considered more explicitly biologically or even humanly motivated settings \citep{gintisXjtb03, axelrodXe04, mcnamaraXn08, scheuringXjtb09, dingXijmpc09}. For example, internal norms, being a pattern of behavior enforced in part by internal sanctions, such as shame, guilt and loss of self-esteem, were found to provide support for the evolution of altruistic norms, and moreover, via a gene-culture coevolution argument an explanation was provided as to why individually fitness-reducing internal norms are likely to be prosocial rather than socially harmful \citep{gintisXjtb03}. Although mentioned already in Section~\ref{aging}, the study of \citet{mcnamaraXn08} in fact focuses on the coevolution of choosiness, the later relying on cooperativeness being used by other individuals as a choice criterion. In such a setting competition to be more generous than others can emerge, and in this case the evolution of cooperation between unrelated individuals can be driven by a positive feedback between increasing levels of cooperativeness and choosiness. It was shown that, in situations where individuals have the opportunity to engage in repeated pairwise interactions, the evolution of cooperation depends critically on the amount of behavioral variation that is being maintained in the population by processes such as mutation.

Finally, we note that \citet{hatzopoulosXpre08} investigated the evolution of cooperation in a so-called nongrowth dynamic network model with a death-birth dynamics based on tournament selection, \citet{hamiltonimXprsb05} as well as \citet{rankinXev09} studied the coevolution of group structure rather than graph structure in the context of generalized reciprocity, while \citet{fortXepl08} considered evolving heterogeneous games as means to sustain cooperation. Interestingly, in the later study the players had individual payoff elements assigned to them for calculating their final payoffs. Accordingly, within the realm of the proposed coevolutionary rule a player could adopt not only the strategy of the neighbor but also its individual payoff matrix elements. It was found that if starting with a random heterogeneous distribution of payoffs, eventually only a small number of definite payoff matrices remained while the others went `extinct'. On the other hand, if the initial rank of individual payoff elements agreed with those constituting social dilemma games, the prevailing one was found to be the stag-hunt game. Considering the latter result in the light of findings reported within the evolving adoption rules model by \citet{szaboXepl09}, it is possible to raise the question if coevolutionary rules as a selection mechanism can spontaneously drive the system into a state where mutual cooperation ensures the maximal average payoff. Indeed, further studies are necessary to clarify this issue. In sum, there are few boundaries to imagination when considering what coevolutionary rules might affect, and certainly, it seems like all facets of existence can be brought into consideration.

\section{Conclusions and outlook}
\label{final}

As we hope the above mini review on coevolutionary games clearly shows, coevolution is certainly a promising concept to follow, as it constitutes the most natural upgrade of evolutionary games in the sense that not only do the strategies evolve in time, but so does the environment, and indeed many other factors that in turn affect back the outcome of the evolution of strategies. Some of these coevolutionary processes are of a finite duration, and thus on their own do not necessarily affect the outcome of evolutionary games but do this only indirectly due to the environment that they produce, while others are lasting, introducing dynamical alterations that affect the evolution of cooperation on a continuous basis. In the future, it should be of interest to further elaborate on the question whether coevolution itself may promote cooperation by introducing dynamical mechanisms, or if mainly the final outcome of a coevolutionary process, if it exists, is the one vital for the sustenance of cooperation. Often, however, it is the interplay of both that facilitates the promotion of cooperation, as it was already shown in some of the works. It should also be considered which coevolutionary processes end sooner or later, and which are those that at least in principle should last forever. For example, the growth of a city can be considered something that has a finite duration due to environmental constrains, while aging, on the other hand, is a natural ingredient of every living organism, and as such it should only make sense to consider it evolving permanently.

Although social dilemmas may emerge at different levels of human and animal interactions, their occurrence is by no means limited to these examples. The applicability of the concept of evolutionary games extends across the whole of social and natural sciences, with examples ranging from the RNA virus \citep{turnerXn99}, ATP-Producing Pathways \citep{pfeifferXs01} and biochemical systems \citep{frickXn03, pfeifferXtbs05, chettaouiXbs07, schusterXjbp08}, to traffic congestion \citep{helbingXacs05, percXnjp07} and climate change \citep{milinskiXpnas06, pfeifferXn06, milinskiXpnas08}, to name but a few. In this sense coevolutionary rules should be applied to evolutionary games in the broadest possible sense, with specially adapted motivation fitting to the research avenue of the main evolutionary process. Moreover, while focusing predominantly on resolving social dilemmas, coevolutionary rules have thus far not been considered for many other game types, as for example the public goods game, the ultimatum game or the rock-scissors-paper game. These gaps should be interesting to fill as well, in particular when striving towards universal concepts underlying cooperation in the broadest possible sense.

\section*{Acknowledgments}
The authors acknowledge support from the Slovenian Research Agency (grant Z1-2032), the Hungarian National Research Fund (grant K-73449), the Bolyai Research Grant, and the Slovene-Hungarian bilateral incentive (grant BI-HU/09-10-001).

\bibliographystyle{elsarticle-harv}

\end{document}